\documentclass[12pt]{iopart}
\usepackage{mathptmx}
\usepackage{graphicx}
\usepackage{times}
\usepackage{amssymb}
\usepackage{url}
\usepackage{subfig}
\usepackage{hyperref}
\usepackage{algorithmic}

\usepackage{lscape}

\begin{document}

\title{Quantitative Analysis of Bloggers Collective Behavior Powered by Emotions}

\author{Marija Mitrovi\'c$^{1}$, Georgios Paltoglou$^{2}$ and Bosiljka Tadi\'c$^{1}$}

\address{$^1$Department of Theoretical Physics, Jo\v zef Stefan Institute, 1001 Ljubljana, Slovenia, \\
$^2${Statistical Cybernetics Research Group, School of Computing and Information Technology University of Wolverhampton, Wulfruna Street, Wolverhampton 
WV1 1LY, UK}, \\
\ead{bosiljka.tadic@ijs.si}  }

%\maketitle
\tableofcontents

%%%%%%%%%%%%%%%%%%%%%%%%%%%%%%%%%%%%%%%%%%%%%%%%%%%%%%%%%%%%%%%%

\begin{abstract}
{ Large-scale data resulting from users online interactions provide the ultimate source of information to study emergent social phenomena on the Web. From individual actions of users to observable collective behaviors, different mechanisms involving emotions expressed in the posted text play a role. Here we combine approaches of  statistical physics with machine-learning methods of text analysis to study emergence of the emotional behavior among  Web users. Mapping the high-resolution data from digg.com onto bipartite network of users and their comments onto posted stories, we identify user communities centered around certain popular posts and determine emotional contents of the related comments by the emotion-classifier developed for this type of texts. Applied over different time periods, this framework reveals strong  correlations between the excess of negative emotions and the evolution of communities. We observe avalanches of emotional comments exhibiting  significant self-organized critical behavior and temporal correlations. To explore robustness of these critical states, we design a network automaton model on  realistic network connections and several control parameters, which can be inferred from the dataset.  Dissemination of  emotions by a small fraction of very active users appears to critically tune the collective states.}
\end{abstract}

\pacs{05.65.+b; 89.20.Hh; 07.05.Tb; 89.75.Fb}
\submitto{JSTAT}
%\keywords{\bf Keywords:} user communities | emotion classifier | collective emotional states | networks | self-organized-criticality}

\section{Introduction}
{\it Online social interactions} among users of different Web portals, which are  mediated by the posted material (text on blogs, pictures, movies, etc.), or via direct exchange of messages on friendship networks, represent a prominent way of human communications. It has been recognized recently \cite{kleinberg2008,berners2006} that  the unsupervised online interactions, involving ever larger number of users through the self-organized dynamics may lead to new social phenomena on the Web. Understanding the emergent collective behavior of users thus appears as one of the central topics of the contemporary science of the Web, beside the structure and the security issues \cite{berners2006,cho2009}.

{\it Role of emotions} known in conventional social contacts has been increasingly perceived in the Web-based communications. The empirical analysis of sentiment and mood, and  opinion mining through user-generated textual data are currently developing research fields \cite{bollen2009,norcross2006,thelwall2009c}. Different dimensions of the emotional state, i.e., arousal, valence and dominance of an individual user can be measured  in the laboratory \cite{thelwall2009c}.   The amount of emotions expressed in a written text and transfered from/to  a user   have been studied \cite{pang2008,thelwall2009c}.   On the other hand, methods  are devised to measure group emotional states and public mood, e.g.,  related to a given event \cite{dodds2009,gonzalez2010}.
  However, the emergence of the collective emotional states from the actions of individual users over time is a nonlinear dynamical process,  that has not been well understood.

{\it Physics and computer science research} of the Web  have been independently developing  own methodologies and goals. For instance, large efforts in the computer science are devoted to improve the algorithms to retrieve information and sentiment from written text \cite{ml,pang2008}. Different methods have been developed for social sciences to analyze particular phenomena \cite{brumfiel2009,panzarasa2009,leskovec2007}. Whereas, physics research is chiefly focused on the underlying processes from the perspective of complex dynamical systems \cite{cattuto2009,sano2009,mitrovic2009c}. 
The quantitative approaches are based on the network representations and application of the graph theory (for a recent review see \cite{boccaletti2006}). 

 Here we use  the theory of complex networks and the methods of statistical physics of self-organized dynamical phenomena, which we  combine with  recent developments in computer science focused towards the emotion contents of the text,  and  study the emergence of collective emotional states among Web users.
  This combined research framework  offers new insight into genesis and structure of the collective emotional states. At the same time it introduces a set of quantitative measures and parameters which characterize users behavior and  can be inferred from the information embedded in the original data. For further understanding of the observed collective states, we design a network-automaton model on the realistic network structure, within which we  tune the control parameters of the dynamics and monitor their effects.

The organization of the paper is as follows. In section\ \ref{sec-data} we explain our methodology and the structure collected data needed for the quantitative analysis of this type. In section\ \ref{sec-empirical} we define the quantities necessary to characterize the collective emotional states of users and perform the systematic analysis of the data to determine these quantities.  In section\ \ref{sec-model} the network-automaton model is introduced and its parameters estimated from the empirical data. The results of simulations are presented. Finally, a short summary and conclusions are given in section\ \ref{sec-conclusions}. Some related technical details are given    in the appendix. 

\section{Data Structure and Methodology\label{sec-data}}  
{\it Fine structure of the data} is required for this type of analysis. Specifically,
we consider a large dataset collected from {\slshape digg.com}, described in the Appendix.
Typically, a user posts a story by providing a link to other media and offering a short description. Then all users may read the story as well as already existing comments, and post own comments, dig (approve), or bury (disapprove) the story. Each user has a unique ID. Every action of a user  is registered with high temporal resolution and clearly attributed to the post (main story) and/or to a given comment on that story. Our data also contain full text of all comments.

Developing an {\it emotion classifier}, which is based on  machine-learning methods \cite{engage} and trained on large dataset of blog texts (see more details in the Appendix), we determine the emotional content of each comment in our dataset. In particular, a probability is determined for each text to be classified first as either {\it subjective}, i.e., having emotional content,  or otherwise {\it objective}. Then the subjective texts are further classified for containing either {\it negative}  or {\it positive} emotional content.  Owing to the high resolution of our data, we are able to study quantitatively the temporal evolution of connected events and determine how the emotions expressed in user's comments affect it. 
Here we are  particularly interested in the collective dynamical effects that emerge through the actions of individual users. 
For this purpose we select a subset consisting of {\it popular posts} with large number of comments,  on which we find over 50\% comment-on-comment actions. The dataset, termed discussion-driven-Diggs (ddDiggs), consists of $N_p$=3984 stories on which $N_C$=917708  comments are written by $N_U$=82201 users. 

\begin{figure}[htb]
\centering
\begin{tabular}{cc}
\resizebox{16.4pc}{!}{\includegraphics{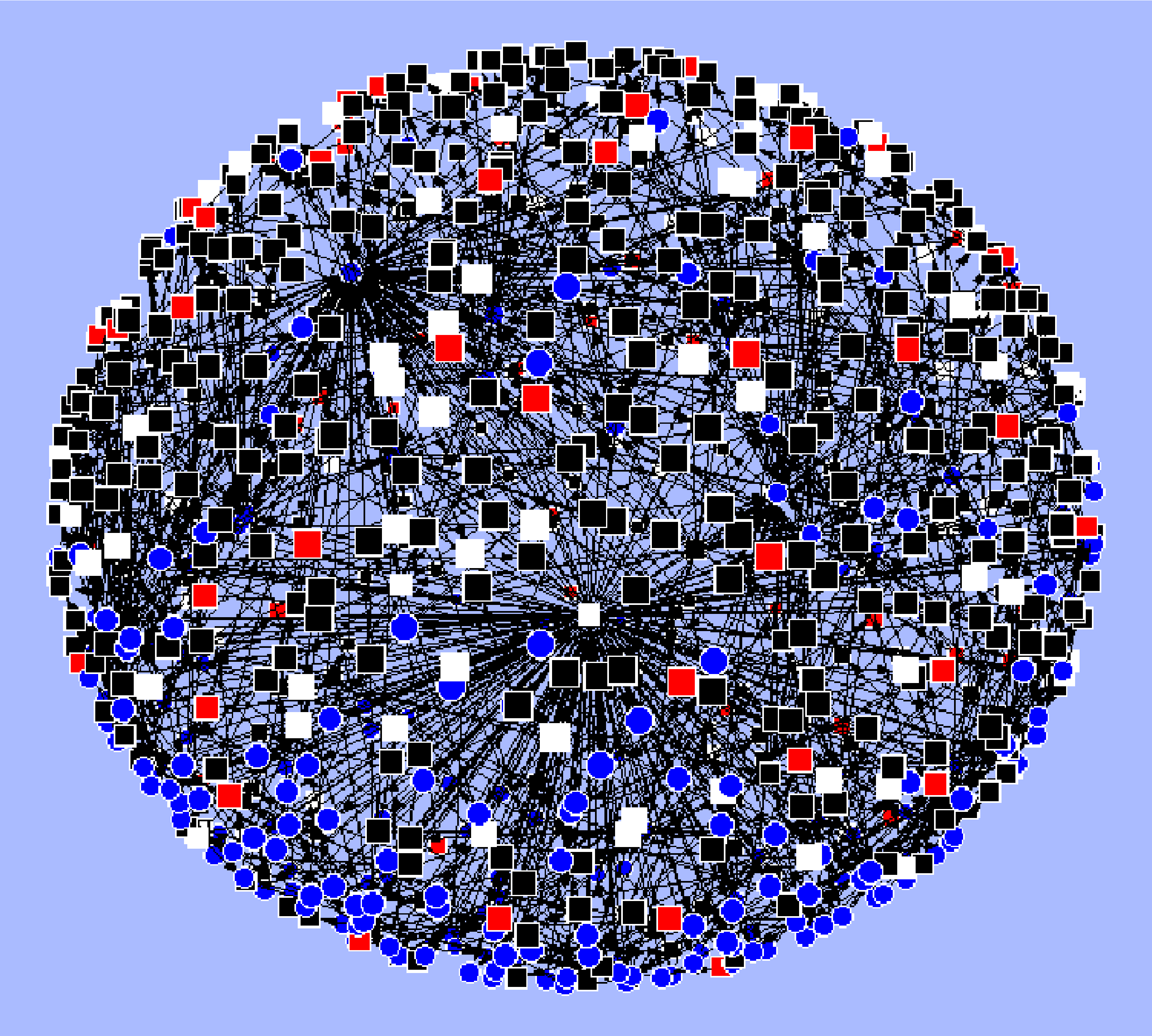}}&
\resizebox{16.4pc}{!}{\includegraphics{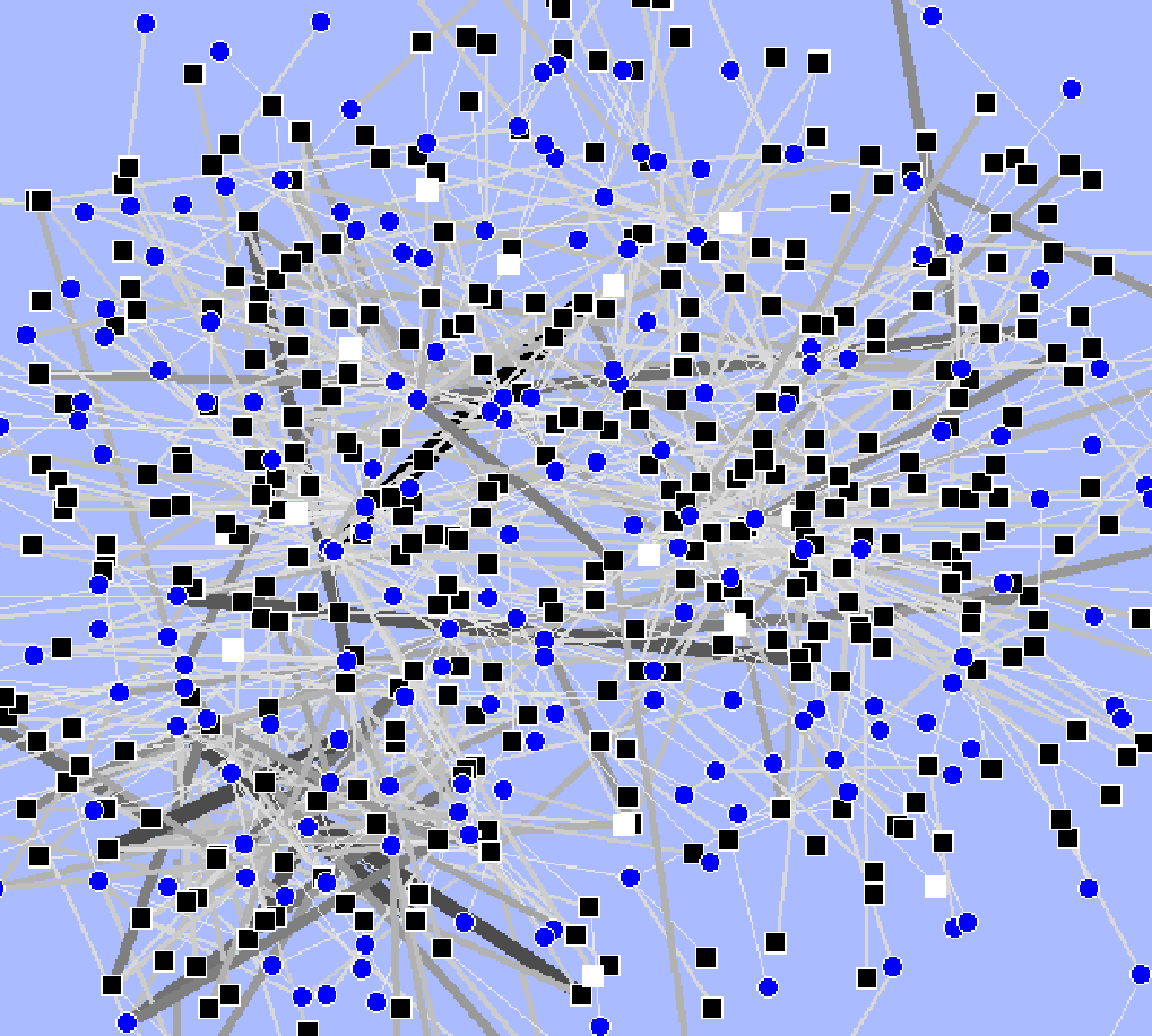}}\\
%\resizebox{20.4pc}{!}{\includegraphics{Figure_1_hr.eps}}\\
\end{tabular}
\caption{(left) One-post bipartite network of users--bullets and comments--squares marked by the emotional contents: red--positive, black--negative, white--neutral. (right) Part of a weighted bipartite network with users (bullets) and posts (squares). The widths of links are given by the number of comments of the user to the post. Color of the post node indicates overall emotional content of all comments on that post.}
\label{fig-bipnets}
\end{figure}

{\it Mapping the data onto a bipartite network} is a first step in our methodology. Two partition nodes are {\it user} nodes and {\it posts-and-comments} nodes, respectively. The resulting network is thus given by $N=N_P+N_C+N_U$ nodes. By definition, a link may occur only between nodes of different partitions, thus these bipartite  networks represent accurately the post-mediated interactions between the Web users. 
(Note that the post-mediated communication makes  the networks of blog users essentially different from the familiar social networks on the Web, such as MySpace or Facebook, where users interact directly with eachother.)
 We keep information about the direction of the actions, specifically, a link $i_P\rightarrow j_U$ indicates that user $j$ reads the post $i$, while $j_U\rightarrow k_c$ indicates that the user $j$ writes the comment $k$. In the data each comment has an ID that clearly attributes it with a given post (original story). The emotional content $e\in [0,-1,+1]$ of the text appears as a {\it property} of each post-and-comment node. Fig.\ \ref{fig-bipnets}(left) shows an example of the accurate data mapping onto directed bipartite network: it represents a network of one popular post from our dataset. Note two types of well connected nodes: the main post (white square visible in the lower part), and a user-hub (circle node visible in upper left part of the network), which indicates a very active user on that post.

Mapping of the entire dataset results in a very large network. For different purposes, however, one can suitably reduce the network size. For instance, a monopartite projection on user-partition can  be made, using the number of common posts per pair of users as a link \cite{mitrovic2009c}. For the purpose of this work, we keep the bipartite representation while we compress the network to obtain a {\it weighted bipartite network} of the size $N=N_P+N_U$, consisting of all popular posts and users attached to them. The weight $W_{ij}$ of a link  is then given by the number of comments of the user $i$ on the post $j$. A part of such network from our ddDiggs data is shown in Fig.\ \ref{fig-bipnets}(right).  These networks exhibit  very rich topology and interesting mixing patterns \cite{MMBT_Springer10,mitrovic2009c} (see also \cite{lambiotte2005,grujic2009} for similar networks constructed from the data of music and  movie users).

Based on the network representation and information about the emotional contents of the texts and the action times, here we perform quantitative analysis of the data. Specifically, we determine:

\begin{itemize}
\item  {\it Community structure} on the weighted bipartite  networks, where a community consists of users and certain posts which are connecting them. Emotional contents of the comments by users in these communities are analyzed;

\item  {\it Temporal patterns of user actions} for each individual user and for the detected user communities; Correlations between the evolution of a community with the emotional contents of the comments is monitored over time;
\item  {\it Avalanches of (emotional) comments}, defined as sequences of comments of a given emotion which are mutually connected over the network and within a small time bin $t_{bin}$=5 minutes. 
\end{itemize}

\section{Empirical Data Analysis\label{sec-empirical}}
 \subsection{User Communities and Emotions}

Time sequence of user activity on all  posts, i.e., the number of comments within a small time bin $t_{bin}=5$ minutes  is followed within the entire time period available in the dataset. Similarly, the time-series of the number of {\it emotional comments}, $n_e(t)$, and the number of {\it negative/positive emotion  comments}, $n_\pm(t)$, is determined. An example is shown in Fig.\ \ref{fig-emots}(top): Zoom of the initial  part of the time-series  is shown, indicating bursts (avalanches) in the number of comments (further analysis of the avalanches is given  below in Figs.\ \ref{fig-popSOCe} and \ref{fig-bipart1P}f).  The occurrence of increased activity over a large period of time  suggests possible formation of a user community around some posts. In this example, the intensive activity with avalanches  of comments lasted  over 2153 hours, followed by reduced activity with sporadic events for another 1076 hours.

Such communities can be accurately identified on  the underlying network by different methods \cite{fortunato2009,rosvall2008,evans2009}. Here we use the method based on the eigenvalue spectral analysis  of the Laplacian operator  \cite{dorogovtsev2003,mitrovic2009b}, which is related to the symmetrical weighted network $\{W_{ij}\}$ as
\begin{equation}
{\cal{L}}_{ij} = \delta _{ij} - \frac{W_{ij}}{\sqrt{\ell_i\ell_j}} \ ; \ell_i=\sum_jW_{ij} \ ,
\label{eq-lap}
\end{equation} 
and $\ell_i$ represents the {\it strength} of  node $i$. As described in detail in \cite{mitrovic2009b}, the existence of  communities in a network is visualized by the branched structure in the scatter-plot of {\it the eigenvectors corresponding to the lowest non-zero eigenvalues of the Laplacian}. 
The situation shown in Fig.\ \ref{fig-bipart1P}a is for the user-projected network of our dataset of popular diggs. In this case $W_{ij}$ represents the number of common posts per pair of users,  users with strengths $\ell _i>100$ are  kept. Branches indicating three large communities are visible. By identifying the nodes' indexes  within a branch, we obtain the list  of users belonging to a community that the branch represents. In order to unravel what posts (and comments) keep a given community together, we perform the spectral  analysis of the weighted bipartite network described above, where the matrix elements $W_{ij}$ represent the number of comments of a user $i$ to post $j$. In this way we identify the list of user- and post-nodes belonging to a community. For each identified community we then select from the original data all comments by the users in that community, together with their time of appearance and the emotional contents.

\subsection{Temporal Patterns of User Behavior}
The actual evolution of an identified community can be retrieved from the data related to it. The procedure---network mapping, community structure finding, and identifying the active users within the community, is repeatedly applied for successive time periods in the past.

\begin{figure}[htb]
\centering
\resizebox{18.8pc}{!}{\includegraphics{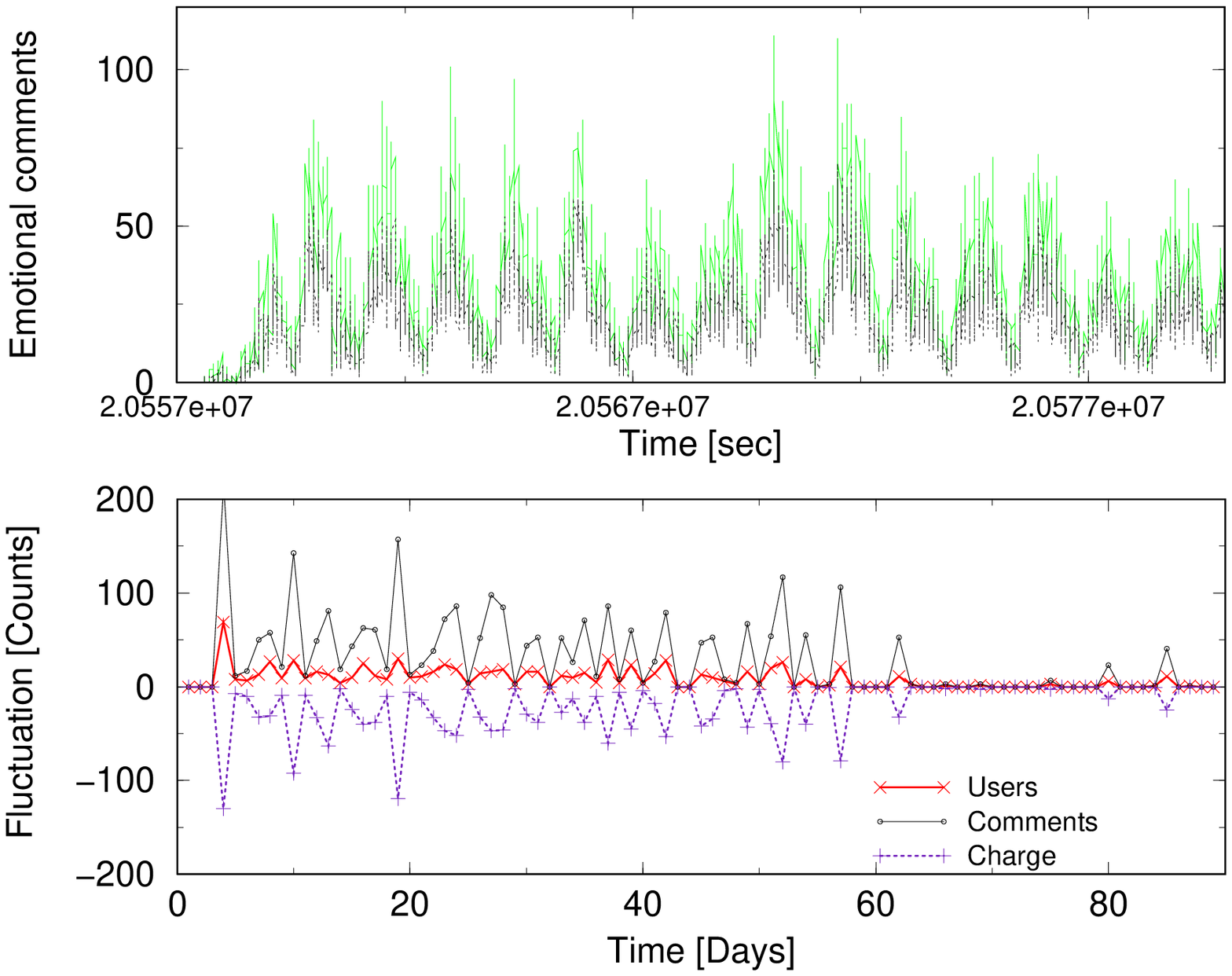}}\\
\caption{(top) Time-series of the emotional comments--pale, and the comments classified as carrying negative emotions--black, in the data of popular digg stories.  (bottom) Temporal fluctuations of the size, the number of comments, and the charge of the emotional comments of an user community on ddDiggs.}
\label{fig-emots}
\end{figure}

 Fluctuations in the number of different users (size) in one of the large communities identified in our ddDiggs dataset is shown in Fig.\ \ref{fig-emots}(bottom) with the time interval of one day.  Shown are also the fluctuations in the number of all comments, and the ``charge'' of emotional comments $Q(t)\equiv n_+(t)-n_-(t)$, where $n_+(t)$ and $n_-(t)$ stand for the number of comments of all users in that community on a given day that are classified as positive and negative, respectively. It is remarkable that, the increase in the number of users is  closely correlated with the excess of the  negative comments (critique) on the posts.  In the supportive material \cite{SI} we give several snapshots of the networks, indicating the evolution stages of this particular community.

\begin{figure*}[!]
\centering
\resizebox{34.4pc}{!}{\includegraphics{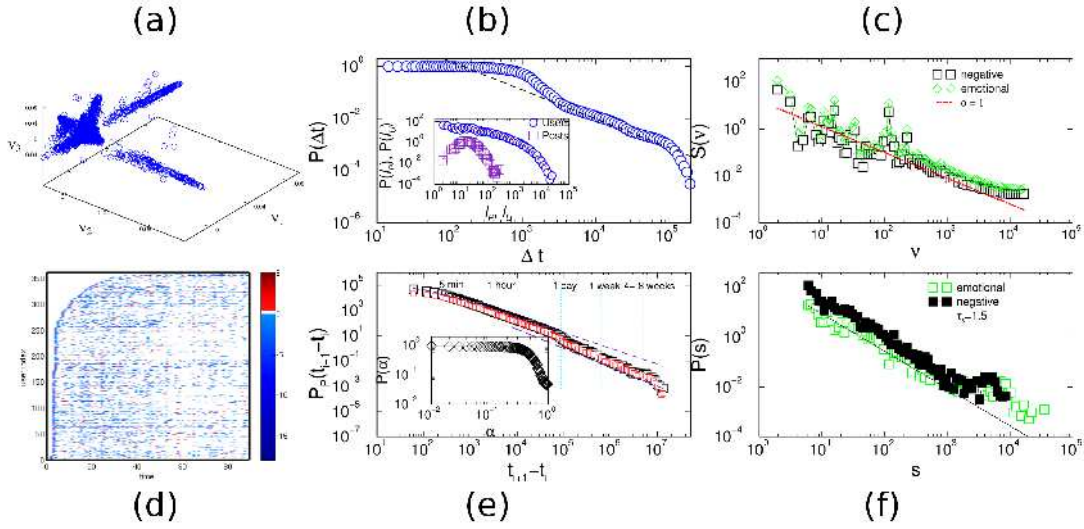}}\\
\caption{(a) Scatter plot of the eigenvectors corresponding to three lowest nonzero eigenvalues of the Laplacian in Eq.\ (\ref{eq-lap}) indicating occurrence of user communities on popular diggs. (b) Distribution of time delay $\Delta t$ between two consecutive user actions, averaged over all users on ddDiggs. Inset: Distribution of strength $\ell_U, \ell_P$ for user and post nodes on the weighted bipartite network, part of which is shown in Fig.\ \ref{fig-bipnets} right. (c) Power-spectrum of the time-series in Fig.\ \ref{fig-emots} top, with all emotional comments--pale, and comments classified as negative--black.     
 (d) Activity pattern of a user community, color indicates charge of all comments by a user within time-bin of 12h; (e) Distribution of the delay 
between the emotional actions of users to a given post, averaged over all posts. Inset:  Distribution of the probability $\alpha$ for a user to post a negative comment.
 (f) Distribution of the size of avalanches with negative--black and all emotional comments--pale, obtained from the same dataset as (c). [Log-binning with the base $b=1.1$ is used for the distributions in figures (b),(c),(e),(f).] }
\label{fig-bipart1P}
\end{figure*}

Users activities on posted text exhibit robust features, which can be characterized by several quantities shown in Figs.\ \ref{fig-bipart1P}(b,c,d,e). Specifically, pattern of a user activity  represents a fractal set along the time axis, with the intervals $\Delta t$ between two consecutive actions obeying a power-law distribution. In Fig.\ \ref{fig-bipart1P}b the histogram $P(\Delta t)$ averaged over all users in ddDiggs is shown. Consequently, the number of comments of a user within a given time bin is varying, that is shown graphically in the color plot in Fig.\ \ref{fig-bipart1P}d. The color code represents the charge of the emotional comments   made by a given user within 12 hours time bin. Different users are marked by the indexes along y-axis and ordered by the time of first appearance in the dataset. Occurrence of the diagonal stripes indicate the activity that involves new users potentially related with the same story. (Note that mutually connected comments are accurately determined  using the network representation, as discussed below). Another power-law dependence is found in the delay time $t_i-t_0$ of comments made by anyone of the users to a given post, measured relative to its posting time $t_0$ \cite{mitrovic2009c,crane2009}. In view of the emotional comments on a given post,  in the present study it is interesting to consider the delay $\delta t\equiv t_{i+1}-t_i$ {\it between two emotional comments}.   
The histograms for the case of  negative (positive) comments from our dataset is given  in Fig.\ \ref{fig-bipart1P}e, averaged over all post in the dataset. 
Both distributions have a power-law tail with the slope $\sim 1.5$ for the delay time in the range $\delta t \in $[24h,8wk] and a smaller slope $\sim 1.25$  for $\delta t \in$[5min,24h], indicated by dashed lines. However, differences and larger frequency of negative comments is found in the domain $\delta t \lesssim$5min. 

Further analysis of the time series  reveals long-range correlations in  the number of emotional comments. In particular, the power-spectrum of the type $S(\nu)\sim 1/\nu$ is found,  both for the number of all emotional comments and the number of comments with negative emotion of the time series from Fig.\ \ref{fig-emots}(top). The  power-spectrum plots are shown in Fig.\ \ref{fig-bipart1P}c. 

\subsection{Structure of the Emergent Critical States}
The observed correlations in the time series are indicative of bursting events, which are familiar to self-organized dynamical systems. In our case an avalanche represents a sequence of comments, i.e.,  a comment triggering more comments  within a small time bin $t_{bin}$, and so on, until the activity eventually stops. In analogy to complex systems as the earthquakes \cite{corral2004} or Barkhausen noise \cite{spasojevic1996}, the avalanches can be readily determined from the measured time-series, like the one shown in Fig.\ \ref{fig-emots}(top). Specifically, putting a baseline on the level of  random noise, an avalanche encloses the connected portion of the signal above the baseline. Thus the {\it size} of an avalanche in our case is given by the number of comments enclosed between two consecutive intersections of the corresponding signal with the baseline. The distribution of sizes of such avalanches is shown in Fig.\ \ref{fig-bipart1P}f, determined from the signal of emotional comments from Fig.\ \ref{fig-emots}(top). A power-law with the slope  $\tau_s\simeq$  1.5 is found over two decades.

The scale-invariance of avalanches is a signature of {\it self-organizing critical (SOC) states}  \cite{dhar2006,jensen1998} in dynamical systems. Typically, a  power-law distribution of the avalanche sizes 
\begin{equation}
P(s)\sim  s^{-\tau_s}exp(-s/s_0) \, 
\label{eq-avalanches}
\end{equation}
 and other quantities pertinent to the dynamics \cite{dhar1990,tadic1996,spasojevic1996} can be measured before a natural cut-off  $s_0$, depending on the system size. The related measures, for instance the distribution of temporal distance between consecutive avalanches, $P(\delta T)$, also exhibits a power-law dependence, as  found in the earthquake dynamics \cite{corral2004,sornette2008}.

\begin{figure}[htb]
\centering
\resizebox{32.0pc}{!}{\includegraphics{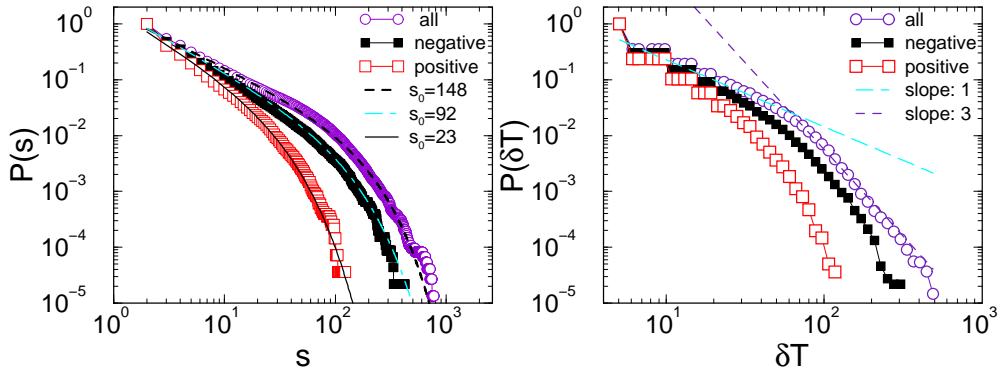}}
\caption{Cumulative distribution of avalanche size $P(s)$, and inactivity time $P(\delta T)$ between avalanches of comments  observed at individual posts on ddDiggs. The cases with comments of positive and negative emotional content are shown separately.}
\label{fig-popSOCe}
\end{figure}

 Here we give evidence that the SOC states may occur in  the events at {\it individual posts} in our dataset.  The results of the cumulative distributions of the avalanche sizes, $P(s)$, averaged over all 3984 posts, are shown in Fig.\ \ref{fig-popSOCe}. The distributions for avalanches of different emotional contents  are fitted with the Eq.\ (\ref{eq-avalanches}) with different exponents ($\tau_s \in[1.0,1.2]$) and cut-offs. On the single-post networks we can also identify the quiescence times {\it between} consecutive  avalanches, the distributions, $P(\delta T)$,  are also shown in Fig.\ \ref{fig-popSOCe}. 
For comparison, the differential distribution of the avalanche sizes in Fig.\ \ref{fig-bipart1P}f, which refers to the simultaneous activity on all posts, shows an excessive number of very large avalanches (supercriticality).
Occurrence of different attractors inherent to the dynamics \cite{tadic1997,tadic1999} or coalescence of simultaneously driven events \cite{corral1999,tadic1996} may result in nonuniversal scaling exponents, which depend on a parameter. 
Relevance  of the conservation laws is still an opened question \cite{bonachela2009}. The situation is even more complex for the dynamics on networks. 
 Nevertheless, the SOC states have been identified in different processes on networks \cite{goh2003,tadic2005,cajueiro2010}. 
In order to understand the origin of the critical states in the empirical  data, and their dependence on the user behavior, in the following we design a cellular-automaton type model on the weighted post--user network, within which we identify the realistic parameters governing the dynamics and vary them.

\section{Modeling Avalanche Dynamics on Popular Posts\label{sec-model}}

The microscopic dynamics on Blogs, i.e., a user posting a comment, triggering more users for their actions, etc., can be formulated in terms of {\it update rules} and {\it constraints}, which affect the course of the process and thus the emergent global states. A minimal set of {\it control parameters} governing the dynamics with the emotional comments is described below and extracted from our empirical data of ddDiggs. Specifically:
\begin{itemize}
\item {\it User delay} $\Delta t$ to posted material, extracted from the data is given by a power-law tailed distribution $P(\Delta t)$ in Fig.\ \ref{fig-bipart1P}b, with the slope $\tau_\Delta \approx$ 1 above the threshold time $\Delta_0\sim$ 300 min;
\item {\it User tendency to post a negative comment}, measured by the probability  $\alpha$,  inferred from the data as a fraction of  negative comments among all comments by a given user. Averaged over all users in the dataset, the distribution $P(\alpha)$ is given in the inset to Fig.\ \ref{fig-bipart1P}e. 
\item {\it Post strength} $\ell_{iP}$ is a topological measure uniquely defined on our weighted bipartite networks as a sum of all weights of its links, i.e., the number of users linked to it with multiplicity of their comments. Thus it  is a measure of attractiveness (relevance) of the posted material. Histograms of the strengths of posts and users in our dataset are given in the inset to Fig.\ \ref{fig-bipart1P}b. 

\item {\it User dissemination probability}  $\lambda$ is a measure of contingency of bloggers' activity. It is deduced from the empirical data as the average fraction of the users who are active more than once/at different posts within a small time bin ($t_{bin}=5$ min). In the model we vary this parameter, as explained below.

\item {\it Network structures} mapped from the real data {\it at various instances of time} underlying the evolution of connected events.
Here we use the weighted bipartite network of the ddDiggs data, Fig.\ \ref{fig-bipnets}(right).
\end{itemize}

Within the network automaton model these parameters are implemented as follows: First, the weighted bipartite network is constructed from the selected data and given time interval. To each post on that network we associate its {\it actual strength} $\ell_{iP}$, and to each user a (quenched) probability $\alpha$ taken from the actual distribution $P(\alpha)$.  A well connected user is selected to start the dynamics by  posting a comment on one of its linked posts. The lists of {\it active users} and {\it exposed posts} are initiated.

\begin{figure}[h]
\centering
\begin{tabular}{cc} 
{\large (a)}&{\large (b)}\\
%\resizebox{28.2pc}{!}{\includegraphics{Figure_5_hr.eps}}\\
\resizebox{18.0pc}{!}{\includegraphics{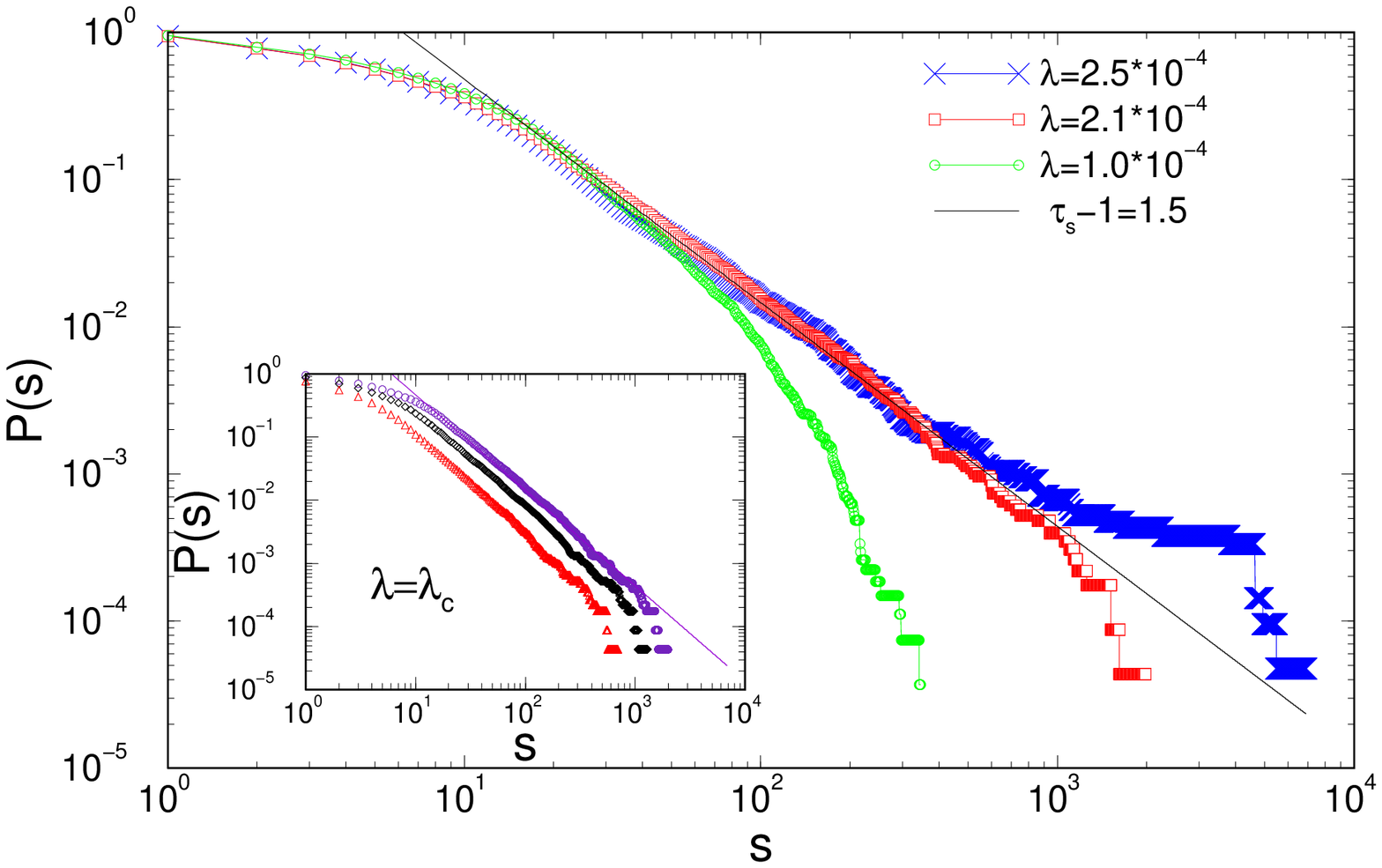}}&
\resizebox{18.0pc}{!}{\includegraphics{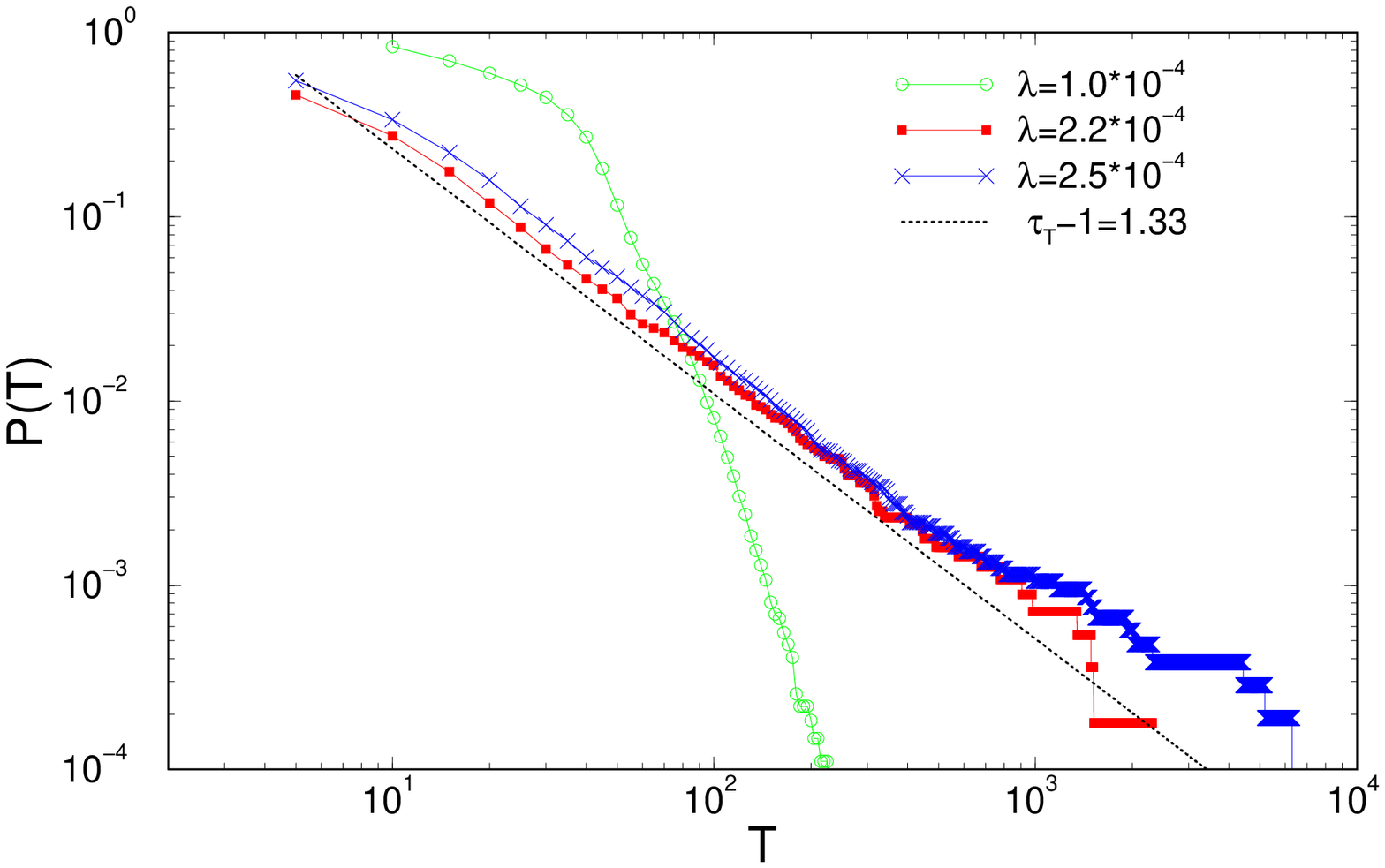}}
\end{tabular}
\caption{Cumulative distributions of the size (a) and duration (b) of avalanches of emotional comments  simulated within the network automaton model for varied dissemination parameter $\lambda$. Fixed parameters for the network structure, posts strengths, and user inclination towards negative comments and delay actions are determined from the ddDiggs dataset, as explained above. Inset: The distribution $P(s)$ for all avalanches, and for avalanches of positive and negative comments for critical value of the dissemination parameter $\lambda=\lambda_c$.
}
\label{fig-socsimul}
\end{figure}

Then at each time step all users linked along the network to the currently exposed posts are prompted for action. A prompted user takes  its delay time $\Delta t$ from the  distribution $P(\Delta t)$ of the actual dataset. Only the users who  got $\Delta t \leq t_{bin}$ are considered as active within this time step and may comment one of the exposed posts along their network links. The posted comment is considered as {\it negative} with the probability $\alpha$ associated with that user, otherwise equal probability applies for the  {\it positive} and  {\it objective} comment. With the  probability $\lambda$ each  active user may  make an additional comment to anyone of its linked (including unexposed) posts.
The post strength is reduced by one with each received comment. Commented post are added to the list of currently exposed posts. In the next time  step the activity starts again from the updated  list of exposed posts, and so on.
Note that the activity can stop when: (a) no user is active, i.e., due to long delay time $\Delta t>t_{bin}$; (b) strength of the targeted post is exhausted; (c) no network links occur between currently active areas.
In the simulations presented here we vary the parameter $\lambda$ while the rest of the parameters are kept at their values inferred from the considered dataset, as described above. The resulting avalanches of all comments and of the positive/negative comments are identified. The distributions of the avalanche  size and duration are shown in Fig.\ \ref{fig-socsimul} for different values of  the dissemination parameter $\lambda$.     

The simulation results, averaged over several initial points, show that the  power-law distributions (\ref{eq-avalanches}) of the avalanche size  occur for the critical value of the dissemination, $\lambda =\lambda_c \sim 2.1\times 10^{-4}$ for this particular dataset. Whereas, varying the parameter $\lambda$ in the simulations appear to have major effects on the bursting process. 
Specifically,  the power-law becomes dominated by the cut-off  for $\lambda  < \lambda_c$, indicating a {\it subcritical} behavior. Conversely, when $\lambda > \lambda_c$, we observe excess of large avalanches, compatible with {\it supercriticality}. The critical behavior at $\lambda =\lambda_c$ has been confirmed by several  other measures. The slopes of the distributions of size and suration, shown in Fig.\ \ref{fig-socsimul} are $\tau_s -1 \approx 1.5$ and $\tau_T\approx 1.33$, respectively.  
The critical behavior  persists, but with changed scaling exponents, when the other control parameters are varied. In particular, assuming the distribution of user-delay as $P(\Delta t)\sim \exp(-\Delta t/T_0)$ leads to the power-law avalanches with the slopes $\tau_s$ and $\tau_T$ depending on the parameter $T_0$.The results are shown in the supporting material \cite{SI}.

\section{Conclusions\label{sec-conclusions}}
We have analyzed a large dataset with discussion-driven comments on digg stories from \textit{digg.com} as a complex dynamical system with emergent collective behavior of users. 
With the appropriate network mappings and using  the methods and theoretical concepts of statistical physics combined with computer science methods for text analysis, we have performed quantitative study of the empirical data to:
\begin{itemize}
\item  Demonstrate how the social communities emerge with users interlinked via their comments over some popular stories; 
\item Reveal that an important part of the driving mechanisms is rooted in the emotional actions of the users, overwhelmed  by negative emotions (critiques); 
\item Show that the bursting events with users' emotional comments exhibit significant self-organization with the critical states.
\end{itemize}
Properties of the emergent collective states can be captured within a network-automaton model, where the real network structure and the parameters native to the studied dataset are taken. Despite several open theoretical problems related with the self-organized criticality on networks, the observed critical states appear to be quite robust when the parameters of users behavior are varied within the model. However, they are prone  to overreaction with supercritical emotional avalanches triggered by a small fraction of very active users, who disseminate activity (and emotions) over different posts.
  Within our approach, the activities and related emotions of every user and of the identified user communities are traced in time and over the emerging  network of their connections. In view of the complex dynamical systems, the statistical indicators of the collective states and the numerical values of the  parameters governing the dynamics of cybercommunities are readily extracted from the empirical data.

\ack {The research leading to these results has received funding from the European 
Community's Seventh Framework Programme FP7-ICT-2008-3 under grant agreement 
n$^o$ 231323 (CYBEREMOTIONS). B.T. also thanks support from the  national program P1-0044 (Slovenia).}

\clearpage

\appendix
\section{}
{\bf Data-Collection:}The Digg data set was collected through the website's publicly available API\footnote{http://apidoc.digg.com/},
which allows programmers to directly access the data stored at its servers, such as stories, comments, user profiles etc.
The data set is comprised of a complete crawl spanning the months February, March and April 2009.
It contains 1,195,808 stories, 1,646,153 individual comments and 877,841 active users. 
More information can be found in \cite{lrec2010}. The data set is freely available for scientific research.

\noindent 
{\bf Emotion Classifier:}The emotion classifier is based on  supervised machine-learning approach, according to which a general inductive process initially learns the characteristics of a class during a training phase, by observing the properties of a number of pre-classified documents, and applies the acquired knowledge to determine the best category for new, unseen documents \cite{ml}.
Specifically, it represents an implementation of the hierarchical Language Model (h-LM) classifier \cite{lms,hierarchical}, according to which a comment is initially classified as objective or subjective and in the latter case, as positive or negative.
The h-LM classifier was trained on the BLOGS06 data set \cite{blog08}, which is a uncompressed 148GB crawl of approximately 100,000 blogs, a subset of which has been annotated by human assessors regarding whether they contain factual information or positive/negative opinions about specific entities, such as people, companies, films, etc. Because the resulting training data set is uneven, the probability thresholds for both classification tasks were optimized on a small subset of humanly annotated Digg comments, in a fashion similar to \cite{engage}.


\section*{References}


\begin{thebibliography}{48}

\bibitem{kleinberg2008}
{Kleinberg} J, 2008 
\newblock {The Convergence of Social and technological Networks}.
\newblock {\em Communications of the ACM}, {\bf 51} 66

\bibitem{berners2006}
Berners-Lee T, Hall W, Hendler J, Shadbolt N, and Weitzner J, 2006
\newblock {Creating a Science of the Web}.
\newblock {\em Science}, {\bf 313}769--771


\bibitem{cho2009}
{Cho} A, 2009
\newblock {Ourselves and Our Interactions: The Ultimate Physics Problem?}
\newblock {\em Science}, {\bf 325} 




\bibitem{bollen2009}
Bollen J, Pepe A, and  Mao H, 2009
\newblock{Modeling public mood and emotion: Twitter sentiment and socio-economic phenomena}.
\newblock{\em arxiv:0911.1583}

\bibitem{norcross2006}
Norcross J C, Guadagnoli E, and Prochaska J O, 2006 
\newblock {Factor structure of the profile of mood states (POMS): Two parallel replications}.
\newblock {\em J. Clinical Psychology}, {\bf 40} 1270-1277
\bibitem{thelwall2009c}
{Thelwall} M, {Buckley} K, {Paltoglou} G, {Cai} D, and {Kappas} A, 2010
\newblock {Sentiment Strenght Detection in Short Informal Text}, \newblock {\em Journal of the Americal Society for Information Science and Technology}, (in press)

\bibitem{pang2008}
Pang B and Lee L, 2008
\newblock {\em Opinion Mining and Sentiment Analysis}.
\newblock Now Publishers Inc.


\bibitem{dodds2009}
{Dodds} P S and {Danforth} C M, 2009
\newblock {Measuring the Happiness of Large-Scale Written Expression: Songs,
  Blogs, and Presidents}.
\newblock {\em Journal of Happiness Studies} 



\bibitem{gonzalez2010}
{Gonzalez-Bailon} S,  {Banchs} R E  and {Kaltenbrunner} A, 2010
\newblock {Emotional Reactions and the Pulse of Public Opinion: Measuring the Impact of Political Events on the Sentiment of Online Discussions}.
\newblock {\em arXiv:1009.4019v1}




\bibitem{ml}
Sebastiani F, 2002 
\newblock Machine learning in automated text categorization.
\newblock {\em ACM Comput. Surv.}, {\bf 34}:1--47



\bibitem{brumfiel2009}
{Brumfiel} G, 2009 
\newblock {Breaking the convention?}
\newblock {\em Nature} {\bf 459} 1050--1051

\bibitem{panzarasa2009}
{Panzarasa} P,  {Opsahl} T  and {Carley} K M, 2009
\newblock Patterns and dynamics of users' behavior and interactions: network
  analysis of and online community.
\newblock {\em Journal of the American Society for Information Science and
  Technology} {\bf 60} 911--932



\bibitem{leskovec2007}
Leskovec J, McGlohon M, Faloutsos Ch, Glance N, and  Hurst M, 2007
\newblock Cascading behavior in large blog graphs.
\newblock {\em In Proceedings of 7th SIAM International Conference on Data Mining (SDM)} p~29406-13


\bibitem{cattuto2009}
Cattuto C, Barrat A, Baldassarri A, Schehr G and Loreto V, 2009
\newblock {Collective dynamics of social annotation}.
\newblock {\em PNAS} {\bf 106}10511--10515



\bibitem{sano2009}
{Sano} Y and {Takayasu} M, 2009
\newblock Macroscopic and microscopic statistical properties observed in blog
  entries.
\newblock {\em arXiv:0906.1744}

\bibitem{mitrovic2009c} 
{Mitrovi\'c} M and {Tadi\'c} B 2010
\newblock Bloggers behavior and emergent communities in blog space.
\newblock {\em Eur. Phys. Journal B} {\bf 73} 293--301



%%%%%%%%%%%%%%%%%%%%%%%%%%%%

\bibitem{boccaletti2006}
{Boccaletti} S, {Latora} L, {Moreno} Y, {Chavez} M  and {Hwang} D U, 2006
\newblock {Complex networks: Structure and dynamics}.
\newblock {\em Physics Reports} {\bf 424} 175--308

\bibitem{engage}
Paltoglou G, Gobron S, Skowron M, Thelwall M and Thalmann D, 2010
\newblock Sentiment analysis of informal textual communication in cyberspace.
\newblock In {\em Proceedings of ENGAGE 2010}, Springer LNCS State-of-the-Art Survey, p~13--23

\bibitem{MMBT_Springer10}Mitrovi\'c M  and Tadi\'c B 2010
\newblock{Emergence and Structure of Cybercommunities.}
\newblock{in Springer Handbook of Networks, Vol. 2, part 2 ``Online Social Networks'', Eds. M.M. Thai and P. Pardalos} (in press)

\bibitem{lambiotte2005}
Lambiotte R and Ausloos M 2005 
\newblock Uncovering collective listening habits and music genres in bipartite
  networks.
\newblock {\em Phys. Rev. E} {\bf 72} 066107

\bibitem{grujic2009}Gruji\'c J, Mitrovi\'c M and Tadi\'c B, 2009
{\em  Mixing patterns and communities on bipartite graphs on web-based social interactions}, IEEEXplore  DOI: 10.1109/ICDSP.2009.5201238,  Digital Signal Processing 16th International Conference, 5-7 July 2009, Santorini, Greece, ISBN: 978-1-4244-3297-4


\bibitem{fortunato2009}
 Fortunato S, 2010
\newblock {Community detection in graphs}.
\newblock {\em Physics Reports} {\bf 486} 75--174

\bibitem{rosvall2008}
{Rosvall} M and {Bergstrom} C T, 2008
\newblock Maps of random walks on complex networks reveal community structure.
\newblock {\em PNAS} {\bf 105} 1118--1123

\bibitem{evans2009}
{Evans} T S and {Lambiotte} R, 2009
\newblock {Line Graphs, Link Partitions and Overlapping Communities}.
\newblock {\em Phys. Rev. E} {\bf 80} 016105



\bibitem{mitrovic2009b}
{Mitrovi{\'c}} M and {Tadi{\'c}} B, 2009
\newblock {Spectral and dynamical properties in classes of sparse networks with
  mesoscopic inhomogeneities}.
\newblock {\em Phys. Rev. E} {\bf 80} 026123


\bibitem{dorogovtsev2003}
{Dorogovtsev} S N, {Goltsev} A V, {Mendes} J F and {Samukhin} A N, 2003
\newblock {Spectra of complex networks}.
\newblock {\em Phys. Rev. E} {\bf 68} 046109

%%%%%%%%%%%%%%%%%%%%%%%%%%%%%%%%%%%%%%%%%%%%%%%%%%%%%%%%%%%%%%%%%%%%%%%%%%

\bibitem{SI}
{Mitrovi{\'c}} M, Paltoglou G and {Tadi{\'c}} B, 2010
\newblock {Supportive Information for: Quantitative Analysis of Bloggers Collective Bahvior Powered by Emotions}.
\newblock {http://www-f1.ijs.si/$\sim$tadic/sup/SI10.pdf}

%%%%%%%%%%%%%%%%%%%%%%%%%%%%%%%%%%%%%%%%%%%%%%%%%%%%%%%%%%%%%%%%%%%%%%%%%%%%%%
\bibitem{crane2009}
{Crane} R, {Schweitzer} F and {Sornette} D, 2010
\newblock {Power Law Signature of Media Exposure in Human Response Waiting
  Time Distributions}.
\newblock {\em Phys. Rev. E} {\bf 81} 056101


%%%%%%%%%%%%%%%%%%%%%%%%%%%%%%%%%%%%%%%%%%%%%%%%%%%%%%%%%%%%%%%%%%%%%%%%%%%
\bibitem{corral2004}
Corral A, 2004
\newblock Long-term clustering, scaling and universality in the temporal
  occurrence of earthquakes.
\newblock {\em Phys. Rev. Lett.} {\bf 92} 108501

\bibitem{sornette2008}
Sornette D, Utkin S and Saichev A, 2008
\newblock Solution of the nonlinear theory and tests of earthquake recurrence times,\newblock{Phys. Rev. E} {\bf 77} 066109

\bibitem{spasojevic1996}Spasojevi\'c D, Bukvi\'c S, Milo\v sevi\'c S and Stanley G, 1996
\newblock{Barkhausen noise: Elementary signals, power laws, and scaling relations},\newblock{Phys. Rev. E} {\bf 54} 2531--46


\bibitem{jensen1998}
Jensen J H, 1998
\newblock {\em Self-organized Criticality. Emergent Complex Behavior in
  Physical and Biological Systems}.
\newblock {Cambridge University Press}
\bibitem{dhar2006}
Dhar D, 2006
\newblock Theoretical studies of self-organized criticality.
\newblock {\em Physica A} {\bf 369} 29--70
\bibitem{dhar1990}
Dhar D, 1990 
\newblock Self-Organized Critical State of Sandpile Automaton Models.
\newblock {\em Phys. Rev. Lett.} {\bf 64} 1613--16
\bibitem{tadic1997}
Tadi\'c B and Dhar D, 1997
\newblock Emergent spatial structures in critical sandpiles.
\newblock {\em Phys. Rev. Lett.} {\bf 79} 1519--22



 \bibitem{tadic1999} Tadi\'c B 1999
\newblock{Dynamic criticality in driven disordered systems: Role of depinning and driving rate in Barkhausen noise}, \newblock{\em  Physica A} {\bf 270} 125--32

\bibitem{corral1999}
Corral A and  Paczuski M, 1999
\newblock Avalanche Merging and Continuous Flow in a Sandpile Model.
\newblock {\em Phys. Rev. Lett.} {\bf 83} 572--575


\bibitem{tadic1996}Tadi\'c B, 1996
 \newblock{Nonuniversal scaling properties of Barkhausen noise.} \newblock{\em Phys. Rev. Lett.} {\bf 77} 3843--6

 \bibitem{bonachela2009}Bonachela J A and Mu\~noz M A, 2009
\newblock{Self-organization without conservation: true or just apparent scale-invariance?}, \newblock{\em J. Stat. Mechanics: Theory and Experiment}, P09009

%%%%%%%%%%%%%%%%%%%SOC-on-networks
\bibitem{goh2003}Goh K I, Lee D S, Kahng B and Kim D, 2003
\newblock{Sandpile on scale-free networks}, \newblock{\em Phys. Rev. Lett.} {\bf 91} 148701-3


\bibitem{tadic2005}Tadi\'c B, Malarz K  and Kulakowski K, 2005
\newblock{Magnetization reversal in spin patterns with complex geometry}, \newblock{\em Phys. Rev. Lett.} {\bf 94} 137204-7

 \bibitem{cajueiro2010}Cajueiro D O and  Andrade R F S, 2010
\newblock{Controlling self-organized criticality in complex networks}, \newblock{\em Eur. Phys. J. B.} 00229-8


%%%%%%%%%%%%%%%%%%%%%%%%%%%%%%%%%%%%5


\bibitem{lrec2010}
Paltoglou G, Thelwall M and Buckely K, 2010
\newblock Online textual communcation annotated with grades of emotion
  strength.
\newblock In {\em Proceedings of the Third International Workshop on EMOTION
  (satellite of LREC): Corpora for research on emotion and affect}, p~25--31


\bibitem{lms}
Peng F, Schuurmans D and Wang S, 2003
\newblock Language and task independent text categorization with simple
  language models.
\newblock In {\em NAACL '03} (Morristown, NJ, USA, 
  Association for Computational Linguistics) p~110--117 

\bibitem{hierarchical}
Pang B and Lee L, 2004
\newblock A sentimental education: sentiment analysis using subjectivity
  summarization based on minimum cuts.
\newblock In {\em ACL '04} (Morristown, NJ, USA) p~271




\bibitem{blog08}
Ounis I, Macdonald C and Soboroff I, 2008
\newblock Overview of the trec-2008 blog trac.
\newblock In {\em The TREC 2008 Proceedings} NIST


\end{thebibliography}
\end{document}